\documentclass[aps,pre,twocolumn,superscriptaddress,floatfix]{revtex4}
\usepackage{graphicx}
\usepackage[dvips,unicode,colorlinks,linkcolor=blue,citecolor=blue,urlcolor=blue]{hyperref}
\usepackage[english]{babel}
\begin{document}
\title{Scroll configurations of carbon nanoribbons}

\author{Alexander V. Savin}
\affiliation{Semenov Institute of Chemical Physics, Russian Academy
of Sciences, Moscow 119991, Russia}

\author{Elena A. Korznikova}
\affiliation{Institute for Metals Superplasticity Problems,
Russian Academy of Science, Khalturin St. 39, Ufa 450001, Russia}

\author{Sergey V. Dmitriev}
\affiliation{Institute for Metals Superplasticity Problems,
Russian Academy of Science, Khalturin St. 39, Ufa 450001, Russia}
\affiliation{Tomsk State University, Lenin Prospekt 36, Tomsk 634050, Russia}

\begin{abstract}
Carbon nanoscroll is a unique topologically open configuration of graphene nanoribbon
possessing outstanding properties and application perspectives due to its morphology.
However molecular dynamics study of nanoscrolls with more than a few coils is limited
by computational power. Here, we propose a simple model
of the molecular chain moving in the plane, allowing to describe the folded and rolled
packaging of long graphene nanoribbons. The model is used to describe a set of possible
stationary states and the low-frequency oscillation modes of isolated single-layer
nanoribbon scrolls as the function of the nanoribbon length. Possible conformational
changes of scrolls due to thermal fluctuations are analyzed and their thermal stability
is examined. Using the full-atomic model, frequency spectrum of thermal vibrations
is calculated for the scroll and compared to that of the flat nanoribbon. It is shown
that the density of phonon states of the scroll differs from the one of the flat
nanoribbon only in the low ($\omega<100$~cm$^{-1}$) and high ($\omega>1450$~cm$^{-1}$)
frequency ranges. Finally, the linear thermal expansion coefficient for the scroll
outer radius is calculated from the long-term dynamics with the help of the developed
planar chain model. The scrolls demonstrate anomalously high coefficient of
thermal expansion and this property can find new applications.
\end{abstract}

\maketitle

\section{Introduction\label{sc1}}

During last decades, various carbon nanostructures have attracted an increasing
attention of researchers, due to their unique electronic, mechanical, and chemical
properties, as well as many potential applications. Numerous studies of
single-layer graphene sheets and graphene nanoribbons (GNR) have started to take
place in recent years \cite{2,3,4,5,6}. Secondary graphene structures
like folds or scrolls can be placed in a separate class of carbon nanomaterials
whose existence is ensured by the action of relatively weak van der Waals bonds
between $sp^2$-bonded monatomic carbon layers. The observation of scrolled graphite
plates under surface rubbing was first reported in 1960 \cite{bs60nature}. The authors
of this work have suggested that the lubricating properties of graphite are due
essentially to the rolling up of packets of layers, which then act like roller
bearings provided a low coefficient of friction. The thickness of the scrolls
was estimated as of order of 100 planes.

The spiral shape and geometric parameters of carbon nanoscrolls (CNS) are determined
by the balance of energy gain due to increase of the number of atoms involved in
van der Waals interactions with the energy loss due to graphene bending.
Several experimental techniques for obtaining and study of CNS have been reported.
\cite{vmk03s,smylkbp07carbon,rabsfdym08cpl,xjfszlflj09nl,cko09carbon,ccl14carbon,zqy11cpl,cbd13nanoscale}
Properties of CNS have also been studied in a series of theoretical
investigations. Electrical, optical and mechanical properties of short CNS have
been described from {\em ab initio} calculations.\cite{pfl05prb,rcg06prb,clg07jpcc}
Mechanical properties  of CNS and various scenarios of their self-assembly have been
described by means of molecular dynamics method.
\cite{bclggb04nl,spcg09apl,mg10nanotechnology,hwf15ss,ppg13jap,wzy15cms,scpg10apl,zl10apl,cxzl11jpcc,psk11acsnano,sgaz13jap,ys13nanoscale}

Mechanical properties and the lowest vibration frequency of long CNS have been
described in the framework of the continuum model of a spiral elastic rod\cite{spcg09apl,scpg10apl,spg10amss,spg11ijf},
where the bending energy of the rod is compensated by the energy gain from the
interaction of adjoining walls.

The resonant oscillation of a CNS near its fundamental frequency might be useful
for molecular loading/release in gene and drug delivery systems.\cite{spcg09apl}
Internal hollow core is one of the main structural features of carbon nanoscrolls. Due to this
cavity the system of coils at low temperatures can serve as an effective storage of hydrogen
atoms,\cite{cbbg07prb,bcbg07cpl} whereas a separate scroll can be used as an ion
channel.\cite{scpg10small}

Under lateral compression hollow cored configuration of the scroll demonstrates
a weak resistance and the core can collapse at a moderate load. This feature opens
the perspective for application of parallel-stacked CNS as an efficient device
sensitive to pressure, which may be used as nano-sized pumps and filters.\cite{spg11ijf,syph13jam}
Simulations buckling and post-critical behavior of CNS under axial compression,
torsion, and bending has revealed the occurrence of kinks and folds.\cite{zhl12jap}

The molecular dynamics technique has proved to be a very powerful tool for
simulation of mechanical properties and deformations mechanisms of CNS.
However, the full-atomic models are very demanding in computational power
making consideration of long-term dynamics of CNS with a large number of coils
almost impossible.

Addressing these challenges, in this paper we propose a simple model of the
planar molecular chain capable of
describing the longitudinal and flexural motion of GNR and allowing the study
of folded and/or scrolled GNR.

In Sec. \ref{sc2} the chain model of the carbon nanoribbon is introduced and the parameters
of the model are fitted to some results in frame of the full-atomic model. In Sec. \ref{sc3}
the chain model is applied to simulate the secondary structures of single layer nanoribbon,
such as folded and scrolled configurations. Then graphene nanoribbon scrolls are analyzed
in more details in Sec. \ref{sc4}. Frequency spectra of the flat nanoribbon and nanoribbon
scroll are calculated in Sec. \ref{sc5} using the full-atomic model. Thermal expansion of
scrolls is analyzed in Sec. \ref{sc6} using the chain model. Section \ref{sc7} concludes
the paper.

\section{Chain model of the carbon nanoribbon \label{sc2}}

\begin{figure}[t,b]
\begin{center}
\includegraphics[angle=0, width=1.\linewidth]{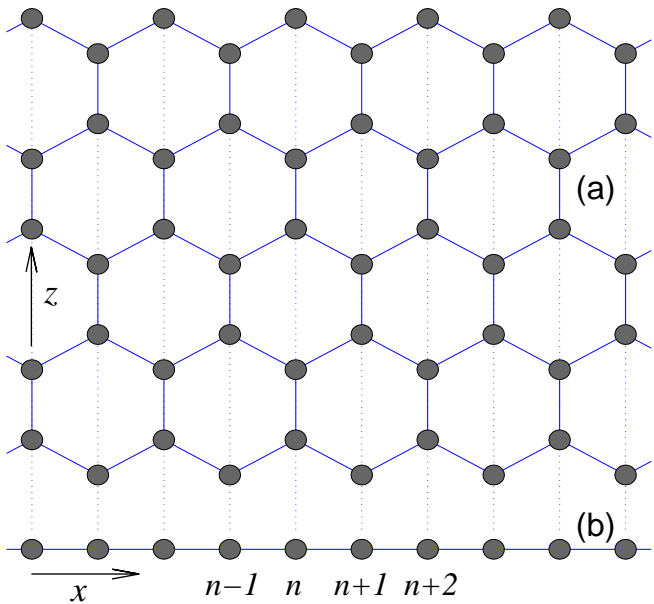}
\end{center}
\caption{
A scheme for constructing the chain model of the carbon nanoribbon.
(a) Full-atomic model of GNR with zigzag orientation; (b) the chain model.
Atomic rows of the nanoribbon oriented along the $z$-axis are numbered by the index $n$.
}
\label{fg01}
\end{figure}

Graphene nanoribbon is a narrow, straight-edged stripe of graphene.
It is well-known that graphene is elastically isotropic material and thus,
its longitudinal and flexural rigidity depend weakly on chirality. For
definiteness, GNR with the zigzag orientation will be considered as shown
in Fig.~\ref{fg01}~(a).

In the flat configuration the nanoribbon is supposed to lay in the $xz$-plane
of the three-dimensional space. The nanoribbon can be described as a periodic
structure with the step $a=r_0\cos(\pi/6)$, where $r_0=0.1418$~nm
is the C-C equilibrium valence bond length. Let us consider such modes of
the nanoribbon motion in which the carbon atoms in the atomic rows parallel
to the $z$-axis move as the rigid units only in the $xy$-plane. Under this
assumption, tensile and flexural GNR dynamics can be described by the chain
of point-wise particles moving in the $xy$-plane. Atomic rows of the nanoribbon
oriented along the $z$-axis are numbered by the index $n$ as shown in
Fig.~\ref{fg01}~(b).

\begin{figure}[t,b]
\begin{center}
\includegraphics[angle=0, width=1\linewidth]{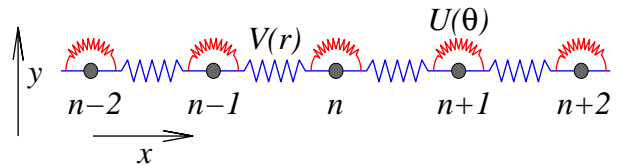}
\end{center}
\caption{
Chain of particles numbered by the index $n$ in the $xy$-plane, modeling
tensile and bending motion of graphene nanoribbon. Potentials $V$ and
$U$ describe the longitudinal and bending stiffness of the chain,
respectively. The potential $W$ (not shown here) describes the weak van
der Waals bonds acting between layers of the chain in the folded or
rolled conformations.
}
\label{fg02}
\end{figure}

The chain model of GNR schematically shown in Fig.~\ref{fg02} can be described by the
following Hamiltonian
\begin{eqnarray}
H=\sum_{n=1}^N\frac12M(\dot{x}^2_n+\dot{y}^2_n)+\sum_{n=1}^{N-1}V(r_n)\nonumber \\
+\sum_{n=2}^{N-1}U(\theta_n)+\sum_{n=1}^{N-4}\sum_{k=n+4}^NW_i(r_{nk}),
\label{f1}
\end{eqnarray}
where $x_n$, $y_n$ are the coordinates of $n$-th particle, $r_n=|{\bf v}_n|$ is the
distance between particles $n$ and $n+1$, with vector
${\bf v}_n=(x_{n+1}-x_n, y_{n+1}-y_n)$ connecting these particles,
$\theta_n$ is the angle between vectors ${\bf v}_n$ and $-{\bf v}_{n-1}$,
$r_{nk}$ is the distance between particles $n$ and $k$ (index $i=1$ if the
difference $k-n$ is an odd number and $i=2$ if $k-n$ is an even number).

The first term in Eq. (\ref{f1}) gives the kinetic energy of the chain with
$M=12m_p$ being the mass of carbon atom ($m_p=1.6603\cdot 10^{-27}$~kg is the proton mass)
and the dot denotes differentiation with respect to time $t$.
The harmonic potential
\begin{equation}
V(r)=\frac12 K(r-a)^2, \label{f2}
\end{equation}
describes the longitudinal stiffness of the chain.

The angular anharmonic potential in Eq. (\ref{f1})
\begin{equation}
U(\theta)=\epsilon[\cos(\theta)+1], \label{f3}
\end{equation}
stands for the flexural rigidity of the chain with
$\cos(\theta_n)=-({\bf v}_{n-1},{\bf v}_n)/r_{n-1}r_n$
being the cosine of the $n$-th "valent"\ angle.

The potential $W_i(r_{nk})$, $(i=1,2)$ in Eq. (\ref{f1}) describes the weak
van der Waals interactions between particles $n$ and $k$, located at the
distance $r_{nk}=\sqrt{(x_k-x_n)^2+(y_k-y_n)^2}$. These interactions, acting
between nanoribbon layers, must be taken into account to describe folded
or scrolled conformations of GNR. Without these interactions the only stable
configuration of a nanoribbon is the flat one.

Parameters of the Hamiltonian $K$ and $\epsilon$ are determined to fit the
dispersion curves of the flat carbon nanoribbon presented in Fig. \ref{fg01}(a)
to the dispersion curves of the straight chain model shown in Fig. \ref{fg02}.
To do so, we consider dynamics of the flat nanoribbon, in line with the chain model,
assuming that the atoms can move only in the $xy$-plane with all atoms in
the rows along $z$-axis displaced equally. Let us denote the coordinates of the
$n$-th atomic row as $x_n$ and $y_n$ and introduce the notation for the vector
${\bf u}_n=(x_n,y_n)$. For the flat nanoribbon the weak van~der~Waals interactions
do not contribute to the dynamics and can be neglected. Then the Hamiltonian for
the nanoribbon can be written in the form
\begin{equation}
H=\sum_{n=-\infty}^{+\infty}\left[ \frac12M(\dot{\bf u}_{n},\dot{\bf u}_n)
+P({\bf u}_{n-1},{\bf u}_{n},{\bf u}_{n+1},{\bf u}_{n+2})
\right].
\label{f4}
\end{equation}
The first term in Eq. (\ref{f4}) describes the kinetic energy, while the second one
stands for the potential energy of interatomic interactions, both per one carbon atom
of $n$-th atomic row, located far from the nanoribbon edges. Thus, the effect of
nanoribbon edges is not taken into account or, in other words, the nanoribbon width
effect is not taken into account.

To describe the carbon-carbon valence interactions let us use a standard set of
molecular dynamics potentials \cite{skh10prb}. The valence bond between two
neighboring carbon atoms $\alpha$ and $\beta$ can be described by the Morse
potential
\begin{equation}
U_1({\bf u}_\alpha,{\bf u}_\beta)=\epsilon_1
\{\exp[-\alpha_0(r-r_0)]-1\}^2,\,\,\,r=|{\bf u}_\alpha-{\bf u}_\beta|, \label{fp1}
\end{equation}
where $\epsilon_1=4.9632$~eV is the valence bond energy and $r_0=0.1418$~nm is the
equilibrium valence bond length. Valence angle deformation energy between three
adjacent carbon atoms $\alpha$, $\beta$, and $\gamma$ can be described by the potential
\begin{equation}
U_2({\bf u}_\alpha,{\bf u}_\beta,{\bf u}_\gamma)=\epsilon_2(\cos\varphi-\cos\varphi_0)^2,
\label{fp2}
\end{equation}
where
$\cos\varphi=({\bf u}_\gamma-{\bf u}_\beta,{\bf u}_\alpha-{\bf u}_\beta)/
(|{\bf u}_\gamma-{\bf u}_\beta|\cdot |{\bf u}_\beta-{\bf u}_\alpha|),
$
and $\varphi_0=2\pi/3$ is the equilibrium valent angle. Parameters
$\alpha_0=17.889$ nm$^{-1}$ and $\epsilon_2=1.3143$~eV can be found from the
small amplitude oscillations spectrum of the graphene sheet \cite{sk08epl}.
Valence bonds between four adjacent carbon atoms $\alpha$, $\beta$, $\gamma$, and $\delta$
constitute torsion angles, the potential energy of which can be defined as
\begin{equation}
U_3(\phi)=\epsilon_3(1-\cos\phi), \label{fp3}
\end{equation}
where $\phi$ is the corresponding torsion angle ($\phi=0$ is the equilibrium value
of the angle) and $\epsilon_3=0.499$~eV.

A detailed discussion of the choice of the interatomic potential parameters can be
found in \cite{skh10prb}. The same set of potentials has been successfully used to
simulate the heat transfer along the carbon nanotubes and
nanoribbons \cite{skh09epl,shk09prb} for the analysis of spatially localized
oscillations \cite{sk94apl,sk10epl,sk10prb,ksbdm12JETPLett,kbd13epl,bdz12epl} and
also for the investigation of theoretical strength and post-critical behavior of
deformed graphene \cite{bdzs12prb,dbsk11JETPLett,bdsk12pss,kd14jpd}.

Hamiltonian Eq. (\ref{f4}) generates the following set of the equations of motion
\begin{eqnarray}
&&-M\ddot{\bf u}_n=F_1({\bf u}_{n},{\bf u}_{n+1},{\bf u}_{n+2},{\bf u}_{n+3}) \nonumber \\
&&+F_2({\bf u}_{n-1},{\bf u}_{n},{\bf u}_{n+1},{\bf u}_{n+2})
+F_3({\bf u}_{n-2},{\bf u}_{n-1},{\bf u}_{n},{\bf u}_{n+1})\nonumber \\
&&+F_4({\bf u}_{n-3},{\bf u}_{n-2},{\bf u}_{n-1},{\bf u}_{n}), \label{f5}
\end{eqnarray}
where vector-function
$$
F_k=\frac{\partial}{\partial {\bf u}_k} P({\bf u}_{1},{\bf u}_{2},{\bf u}_{3},{\bf u}_{4}),
~~k=1,2,3,4.
$$

It is convenient to use the relative coordinates of atoms
${\bf w}_n(t)={\bf u}_n(t)-{\bf u}_n^0$, where ${\bf u}_n^0$ are the
equilibrium coordinates.
For the analysis of small-amplitude vibrations ($|{\bf w}_n|\ll r_0$)
we use the following linearized equations of motion
\begin{eqnarray}
-M\ddot{\bf w}_n=B_1{\bf w}_n+B_2({\bf w}_{n-1}+{\bf w}_{n+1})\nonumber \\
+B_3({\bf w}_{n-2}+{\bf w}_{n+2})+B_4({\bf w}_{n-3}+{\bf w}_{n+3}),
\label{f6}
\end{eqnarray}
where matrices $B_1=F_{11}+F_{22}+F_{33}+F_{44}$, $B_2=F_{12}+F_{23}+F_{34}$,
$B_3=F_{13}+F_{24}$, $B_4=F_{14}$, and matrix
$$
F_{kl}=\frac{\partial}{\partial {\bf w}_k \partial {\bf w}_l}
P({\bf 0},{\bf 0},{\bf 0},{\bf 0}).
$$

We seek for the solution of linear system Eq.~(\ref{f6}) in the form of the wave
\begin{equation}
{\bf w}_n(t)={\bf A}\exp[i(qn-\omega t)],
\label{f7}
\end{equation}
where $\omega$ is the frequency of the wave, ${\bf A}$ is the amplitude vector,
$q\in [0,\pi]$ is the dimensionless wavenumber. Substituting Eq. (\ref{f7})
into the linear system Eq. (\ref{f6}) we obtain the dispersion relation
\begin{equation}
|B_1+2\cos(q)B_2+2\cos(2q)B_3+2\cos(3q)B_4-\omega^2E|=0,
\label{f8}
\end{equation}
where $E$ is the unity matrix.
\begin{figure}[t,b]
\begin{center}
\includegraphics[angle=0, width=1\linewidth]{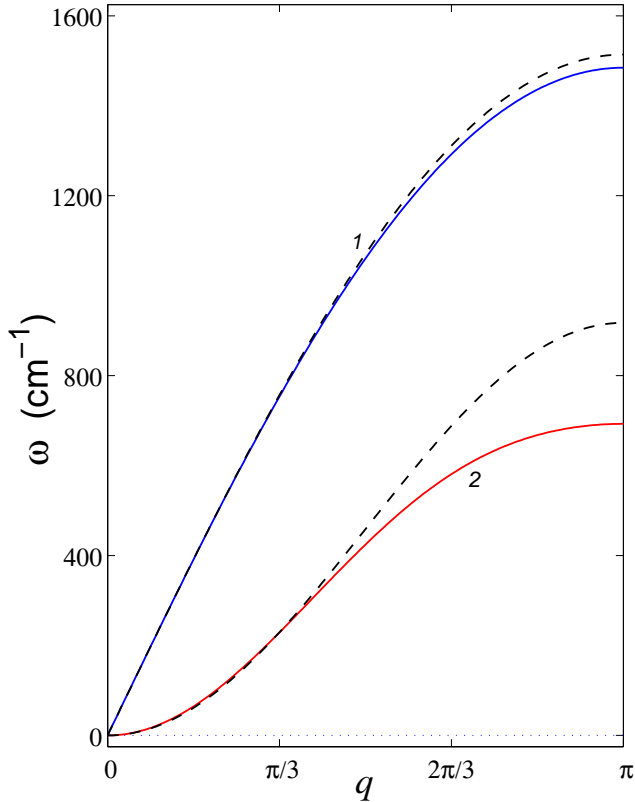}
\end{center}
\caption{
Dispersion curves of the wide carbon nanoribbon. Curve 1 stands for the longitudinal
(tension-compression) and curve 2 for the transversal (bending) phonons.
Solid lines correspond to the full-atomic model, while dashed lines to the chain model
with appropriately chosen parameters.}
\label{fg03}
\end{figure}

The dispersion relation Eq. (\ref{f8}) is the second order polynomial with respect
to the squared frequency $\omega^2$. The corresponding dispersion relation has
two branches $ 0\le\omega_y(q)\le\omega_x(q)$ as plotted in Fig. \ref{fg03} by the
solid lines. The low-frequency branch $\omega=\omega_y(q)$ describes the dispersion of the
transverse plane waves when lattice nodes leave the nanoribbon plane and move along
$y$ axis (bending nanoribbon vibrations). The high-frequency branch $\omega=\omega_z(q)$
represents the dispersion of the longitudinal plane waves when the nodes move along
$x$-axis (longitudinal nanoribbon vibrations).

Velocity of long-wavelength plane phonons coresponds to the velocity of sound which is
$$
v_x=a\lim_{q\rightarrow 0}\omega_x(q)/q=17510~\mbox{m/s}
$$
for the longitudinal phonons and
$$
v_y=a\lim_{q\rightarrow 0}\omega_y(q)/q=0
$$
for bending phonons.

Similarly, we can obtain the dispersion curves for the chain model. In this case,
potential energy in the Hamiltonian Eq. (\ref{f4}) is defined as
$$
P({\bf u}_{n-1},{\bf u}_{n},{\bf u}_{n+1})=V(r_n)+U(\theta_n).
$$
Potential parameters $K$ of Eq. (\ref{f2}) and $\epsilon$ of Eq. (\ref{f3}) should be chosen
in a way to achieve the best fit of the dispersion curves obtained for the full-atomic model.
We are interested in the long-wavelength modes of the nanoribbon motion and thus, the best fit
should be achieved in the range of small values of $q$. The choice
\begin{equation}
K=405~\mbox{N/m},~~~\epsilon=3.50~\mbox{eV},
\label{f9}
\end{equation}
assures the coincidence of the longitudinal and flexural rigidity of the chain model and
the nanoribbon, as can be seen in Fig. \ref{fg03}, where the result for the chain model
is plotted by the dashed lines.

\begin{figure}[t,b,p]
\begin{center}
\includegraphics[angle=0, width=1\linewidth]{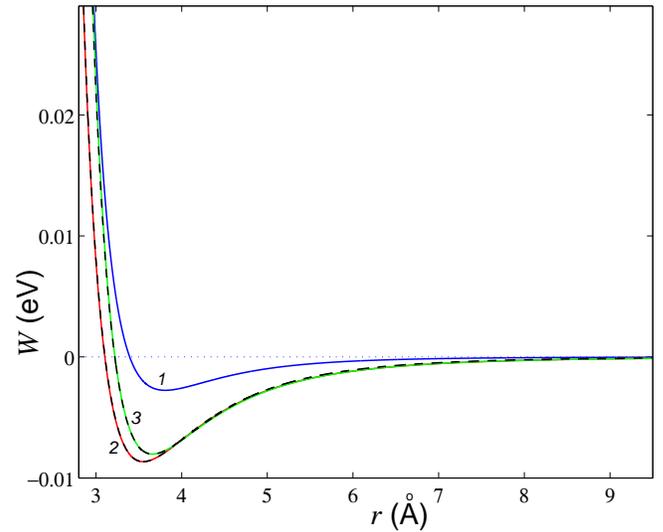}
\end{center}
\caption{
The pairwise Lennard-Jones potential $W_0(r)$ describing the van der Waals interaction
between two carbon atoms (curve 1). Solid lines 2 and 3 show the potentials $W_1(r)$
and $W_2(r)$ (see in the text), while their approximation by the modified Lennard-Jones
potentials Eq. (\ref{f13}) are shown by the dashed lines 2 and 3.
}
\label{fg04}
\end{figure}

The van der Waals interactions between carbon atoms are described by the pairwise
Lennard-Jones potential
\begin{equation}
W_{0}(r)=4\varepsilon[(\sigma/r)^{12}-(\sigma/r)^6],
\label{f10}
\end{equation}
where $r$ is the distance between two carbon atoms. The parameters of the Lennard-Jones
potential $\varepsilon=0.002757$~eV and $\sigma=0.3393$~nm were fitted to reproduce
the interlayer binding energy \cite{p22}, interlayer spacing \cite{p16,p17}, and $c$-axis
compressibility \cite{p18} of graphite. The potential Eq. (\ref{f10}) with these parameters
is shown in Fig. \ref{fg04} by the curve 1.

The long range interaction between the chain nodes $n$ and $k$ is described
by the van der Waals interactions of the atoms belonging to $n$-th and $k$-th
atomic row of the nanoribbon. Therefore, the interaction energy can be expressed as
\begin{eqnarray}
\label{f11}
W_1(r)&=&\sum_{j=-\infty}^{\infty}[W_{0}(r_{j,1})+W_{0}(r_{j,2})],\\
      && r_{j,1}=[r^2+(-0.5r_0+3jr_0)^2]^{1/2}, \nonumber\\
      && r_{j,2}=[r^2+(-1.5r_0+3jr_0)^2]^{1/2}, \nonumber
\end{eqnarray}
if the difference $k-n$ is an odd number and
\begin{eqnarray}
\label{f12}
W_2(r)&=&\sum_{j=-\infty}^{\infty}[W_{0}(r_{j,3})+W_{0}(r_{j,4})],\\
      && r_{j,3}=[r^2+(3jr_0)^2]^{1/2}, \nonumber\\
      && r_{j,4}=[r^2+(-r_0+3jr_0)^2]^{1/2}, \nonumber
\end{eqnarray}
if $k-n$ is an even number.

Interaction potentials Eq.~(\ref{f11}) and
Eq.~(\ref{f12}) are well approximated by the modified Lennard-Jones potential
\begin{eqnarray}
W_i(r)=4\varepsilon_i [(\sigma_i /f(r))^{12}-(\sigma_i/f(r))^6], \nonumber\\
f(r)=r_i(r/r_i)^{\alpha_i},~~~r_i=2^{1/6}\sigma_i.
\label{f13}
\end{eqnarray}
For $i=1$, the modified potential parameters are $\varepsilon_1=0.008652$~eV,
$\sigma_1=0.31636$~nm, and $\alpha_1=0.86$. For $i=2$, parameters are
$\varepsilon_2=0.008029$~eV, $\sigma_2=0.32607$~nm, and $\alpha_2=0.90$.
The interaction potentials of nanoribbon atomic rows Eq.~(\ref{f11}) and
Eq.~(\ref{f12}) together with the corresponding modified Lennard-Jones approximations
Eq.~(\ref{f13}) are shown in Fig.~\ref{fg04} by the solid and dashed lines,
respectively. Practically perfect coincidence of these potentials can be observed.

{\it Summing up}, for GNR shown in Fig.~\ref{fg01}~(a), we have developed the chain
model depicted in Fig.~\ref{fg02} and described by the Hamiltonian Eq.~(\ref{f1})
with the parameters fitted to reproduce the long-wavelength phonon spectrum of GNR
(see Fig.~\ref{fg03}) and the van der Waals interactions acting between carbon atoms
in folded or scrolled conformations of the nanoribbon (see Fig.~\ref{fg04}).
The chain model describes only such modes of GNR deformation in which the atomic
rows parallel to $z$-axis move as rigid units only in the $xy$-plane but not
in $z$-direction. The nanoribbon width effect is not taken into account.

\section{Secondary structures of single layer nanoribbon \label{sc3}}

To find stable structures of a one-layer carbon nanoribbon the following
minimization problem should be considered
\begin{eqnarray}
E=\sum_{n=1}^{N-1}V(r_n)+\sum_{n=2}^{N-1}U(\theta_n) \nonumber\\
+\sum_{n=1}^{N-4}\sum_{k=n+4}^NW_i(r_{nk})\rightarrow \min: \{{\bf u}_n\}_{n=1}^N,
\label{f14}
\end{eqnarray}
where the minimization of potential energy of the chain model having $N$ nodes
is performed with respect to the coordinates of the nodes ${\bf u}_n=(x_{n},y_{n})$,
$n=1,2,...,N$. The nanoribbon length is defined by the number of nodes, $L=(N-1)a$.

The energy minimization was carried out numerically using the conjugate gradient
method. In order to check the stability of the resulting stationary configuration
$\{ {\bf u}_n^0\}_{n=1}^N$ we calculate the eigenvalues of the $2N\times 2N$
matrix of the second derivatives
\begin{equation}
B=\left(\left. \frac{\partial E}{\partial u_{n,i}\partial u_{k,j}}
\right|_{\{ {\bf u}_m^0\}_{m=1}^N}\right)_{n=1,i=1,k=1,j=1}^{~~N,~~2,~~N,~~2}.
\label{f15}
\end{equation}
The stationary chain configuration is stable only if all eigenvalues of a
symmetric matrix $B$  are non-negative: $\lambda_i\ge 0$, $i=1,2,...,2N$.
Note that for stable configuration  the first three eigenvalues are always
zero $\lambda_1=\lambda_2=\lambda_3=0$. These eigenvalues correspond to
the rigid motion of the chain in the $xy$-plane, with two translational and one
rotational degrees of freedom. The remaining positive eigenvalues
$\lambda_i>0$ correspond to the oscillation eigenmodes with frequencies
$\omega_i=\sqrt{\lambda_i/M},~i=4,...,2N$.

\begin{figure}[t,b]
\begin{center}
\includegraphics[angle=0, width=1\linewidth]{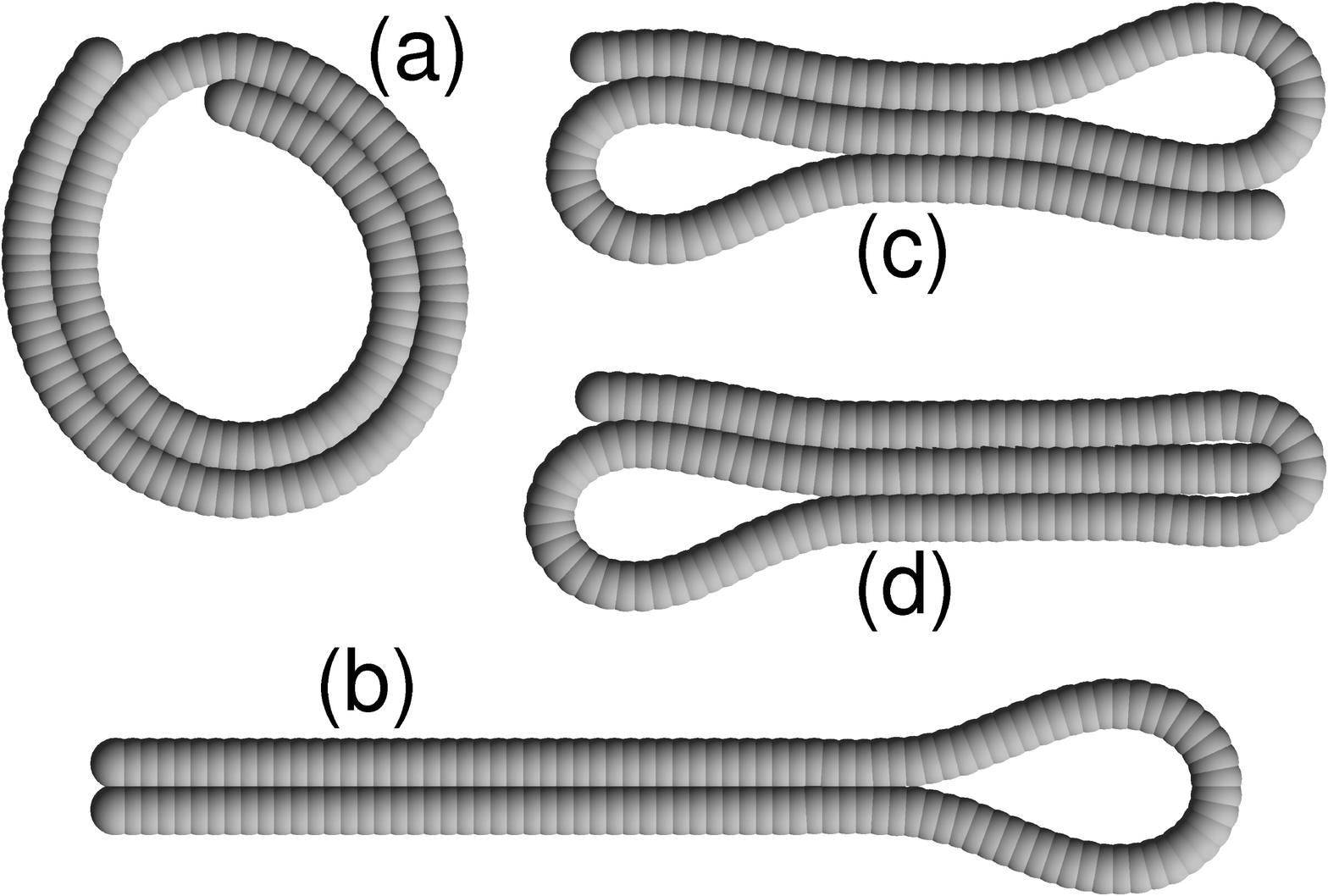}
\end{center}
\caption{
Stable stationary conformations of the chain model with $N=140$ nodes representing
the nanoribbon of length $L=17.070$~nm: (a) rolled, (b) double-folded, (c) triple-folded,
and (d) rolled-collapsed. Not shown here is the straight stable configuration
(flat nanoribbon).
}
\label{fg05}
\end{figure}

Stationary structure of the chain depends on the initial configuration
used to solve the minimization problem Eq. (\ref{f14}). Changing the initial
configuration, a variety of stable configurations can be found. Linear chain
configuration, representing flat nanoribbon, is always stable. The weak
van der Waals interactions between nodes give rise to the existence of other,
more advantageous in energy, stationary states of the chain in the two-dimensional
space. As exemplified in Fig.~\ref{fg05}, the chain consisting of $N=140$ nodes
and having length $L=(N-1)a=17.070$~nm, in addition to the flat state, can be stable in (a) rolled,
(b) double-folded, (c) triple-folded, and (d) rolled-collapsed conformations.
Potential energy per node, $E_0=E/N$, is used to compare the energy of different
chain conformations. In case of $N=140$, the flat structure has
$E_0=-0.00453$~eV. For the other forms one has: $E_0=-0.01395$~eV for the rolled
state, $-0.01214$~eV for the double-folded, $-0.00352$~eV for the triple-folded, and
$-0.00662$~eV for the rolled-collapsed state. To understand these figures,
one should keep in mind that formation of van der Waals bonds lowers the
structure total energy, while the large curvature regions increase the energy.
The rolled packing is the most energetically favorable among the studied
conformations of the nanoribbon. All the studied non-flat structures have
energy lower than the flat one, except for the triple-folded one. This is
explained by the fact that the triple-folded structure possesses two loops
with large curvature having no van der Waals bonds, and such loops have
relatively large energy.
\begin{figure}[t,b]
\begin{center}
\includegraphics[angle=0, width=1\linewidth]{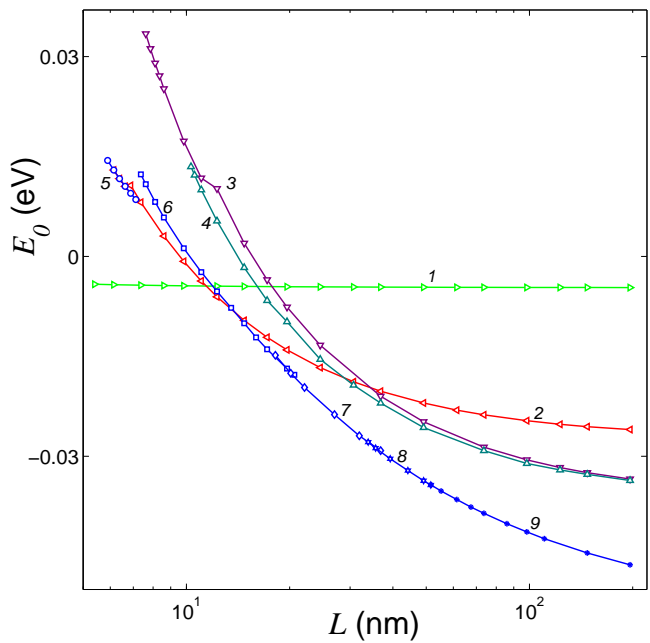}
\end{center}
\caption{
Potential energy per node, $E_0=E/N$, as the function of the chain length $L$
for flat nanoribbon (curve 1), double-folded (curve 2), triple-folded (curve 3),
rolled-collapsed (curve 4), and for the rolled structures with one-coil (curve 5),
two-coil (curve 6), three coil (curve 7), four coil (curve 8) and more coiled
(curve 9) structures of the nanoribbon.
}
\label{fg06}
\end{figure}

The dependence of the normalized energy $E_0$ for different stationary
nanoribbon packages  on its length $L$  is shown in Fig.~\ref{fg06}.
In the range $L<5.77$~nm the planar structure is the only stable
configuration of the nanoribbon. For $L\ge5.77$~nm, stable rolled structures
exist. Chains with $L\ge 6.02$~nm ($L\ge 10.19$~nm) can support stable double-folded
(triple-folded) structures. The rolled-collapsed structure requires
the chain length $L\ge 10.19$~nm.

The flat nanoribbon has the lowest energy for $L< 10.93$~nm. For the
nanoribbon length in the range $10.93 \le L<13.39$~nm the lowest energy
is observed for the double-folded configuration and for $L\ge 13.39$~nm
the most energetically favorable is the rolled structure (nanoribbon scroll).

\begin{figure}[t,b]
\begin{center}
\includegraphics[angle=0, width=1\linewidth]{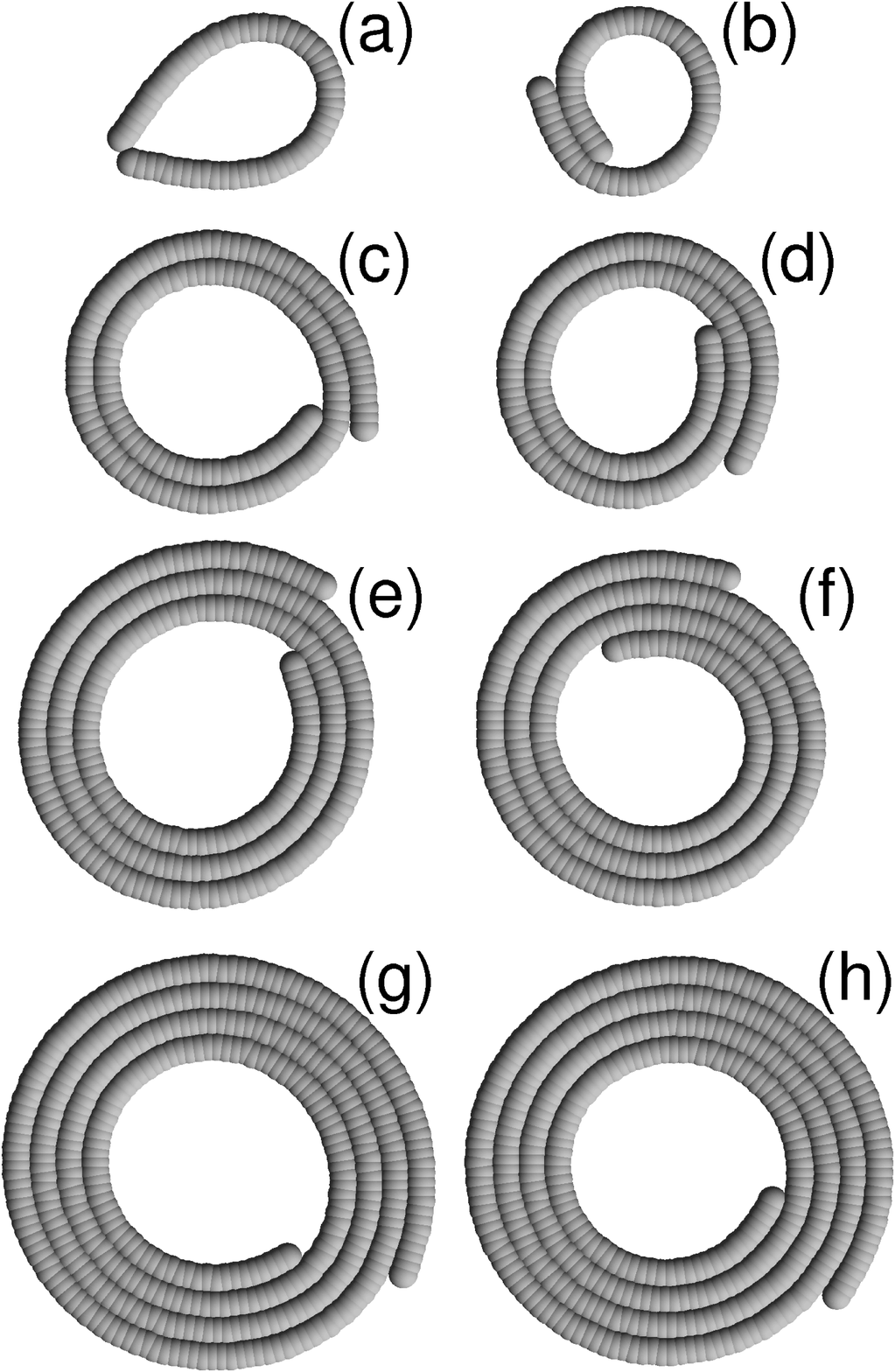}
\end{center}
\caption{
Examples of the stable, equilibrium scroll structures of the nanoribbon:
(a) single coil scroll (nanoribbon length $L=7.00$~nm, energy per node $E_0=0.00881$~eV,
number of coils $N_c=0.96$, inner and outer radii $R_1=R_2=1.11$~nm);
(b) double coil scroll ($L=7.25$~nm,
$E_0=0.0123$~eV, $N_c=1.19$, $R_1=0.97$~nm, $R_2=1.021$~nm);
(c) double coil scroll ($L=19.53$~nm, $E_0=-0.01684$~eV, $N_c=1.98$,
$R_1=1.40$~nm, $R_2=1.73$~nm);
(d) three coil scroll ($L=19.53$~nm, $E_0=-0.01692$~eV, $N_c=2.19$, $R_1=1.24$~nm, $R_2=1.63$~nm);
(e) three coil scroll ($L=34.26$~nm, $E_0=-0.02815$~eV, $N_c=2.96$, $R_1=1.51$~nm, $R_2=2.17$~nm);
(f) four coil scroll ($L=34.26$~nm, $E_0=-0.02816$~eV, $N_c=3.14$, $R_1=1.38$~nm, $R_2=2.10$~nm);
(g) four coil scroll ($L=51.45$~nm, $E_0=-0.03432$~eV, $N_c=3.94$, $R_1=1.58$~nm, $R_2=2.57$~nm);
(h) five coil scroll ($L=51.45$~nm, $E_0=-0.03429$~eV, $N_c=4.05$, $R_1=1.51$~nm, $R_2=2.54$~nm).
}
\label{fg07}
\end{figure}

\section{Graphene nanoribbon scrolls \label{sc4}}

In the preceding Section it was shown that GNR having length $L\ge 13.39$~nm
have the lowest energy in the rolled conformation among the other studied
configurations. That is why, here we focus on the study of nanoribbon scrolls.
The cross-sectional view of the minimum energy scroll structures for nanoribbons
of increasing length can be seen in Fig.~\ref{fg07}. The scroll cross-section
appears in the form of the truncated Archimedes spiral always having inner
cavity. The scroll structure is determined by the balance of energy gain caused
by increasing the number of atoms having van der Waals bonds with the others
and the energy loss due to the increase of nanoribbon curvature.

The center of mass can be considered as the center of the scroll:
$$
{\bf u}_0=\frac1N\sum_{n=1}^N{\bf u}_n^0,
$$
where ${\bf u}_n^0=(x_n^0,y_n^0)$ is the two-dimensional radius-vectors of
the $n$-th chain node of the equilibrium scroll. In the polar coordinate system it
can be written as
\begin{equation}
x_n^0=x_0+R_n^0\cos(\phi_n^0),~~y_n^0=y_0+R_n^0\sin(\phi_n^0),
\end{equation}
where $R_n^0=\sqrt{(x_n^0-x_0)^2+(y_n^0-y_0)^2}$ and the discrete angle $\phi_n^0$
monotonously increases with increasing node
number $n=1,2,...,N$. The spiral can be characterized by the number of coils
$$
N_c=(\phi_N-\phi_1)/2\pi.
$$
It is also convenient to define the integer number of coils $m=[N_c]+1$,
where $[x]$ is the integer part of $x$. Let us define the inner radius
of scroll by its first coil:
$$
R_1=\frac{1}{n_1}\sum_{n=1}^{n_1} R_n^0,
$$
where $n_1$ is the maximal value of index $n$ wherein $\phi_n<\phi_1+2\pi$.
The outer radius of the scroll can be defined by its last coil as:
$$
R_2=\frac{1}{N-n_2+1}\sum_{n=N-n2}^{N} R_n^0,
$$
where $n_2$ is the minimal value of $n$ where $\phi_n>\phi_N-2\pi$.

The twisting rigidity of the scroll is characterized by the lowest
natural frequency $\omega_1=\sqrt{\lambda_4/M}$. This frequency corresponds
to the periodic twisting/untwisting oscillations of the scroll. In the
approximation of a continuous elastic rod this oscillation motion has been
studied in \cite{spcg09apl,spg10amss}.
\begin{figure}[tb]
\begin{center}
\includegraphics[angle=0, width=1\linewidth]{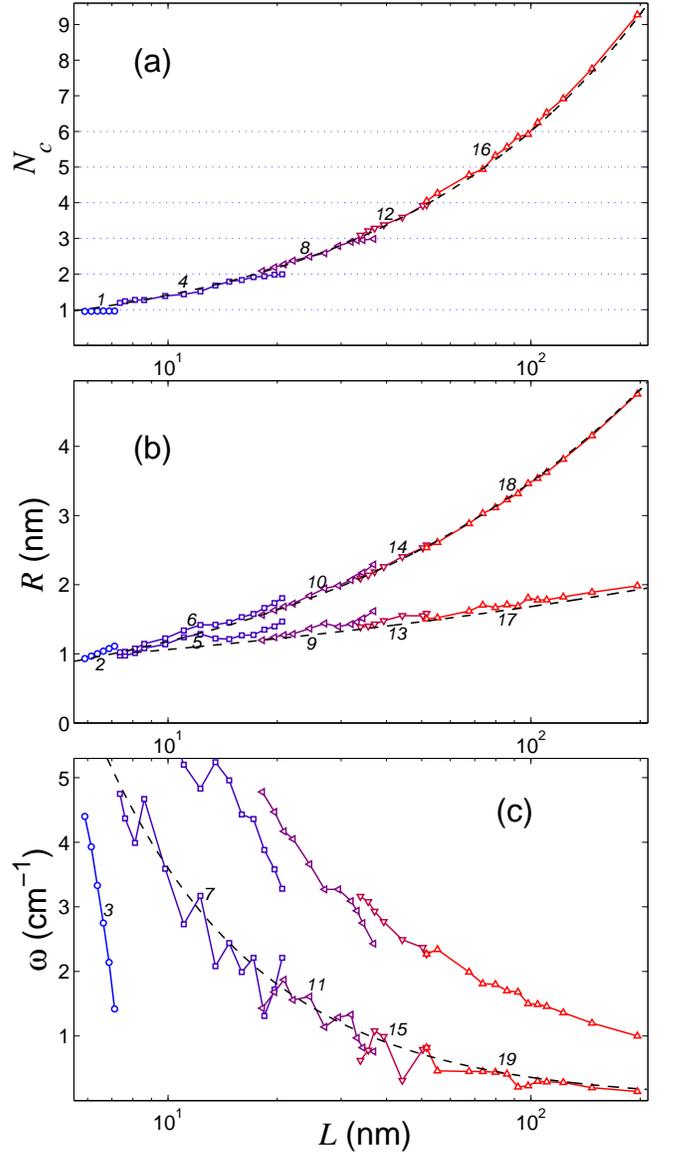}
\end{center}
\caption{
(a) Effect of the nanoribbon length $L$ on (a) the number of coils of the
stationary scroll $N_c$, (b) inner $R_1$ (lower curve) and outer $R_2$
(upper curve) radii of the scroll, (c) the lowest natural vibration frequencies
of the scroll, $\omega_1$ and $\omega_2$.
The curves 1, 2, and 3 correspond to a single-coil
scroll; curves 4-7 to the two-coil scroll; 8-11 to the three-coil scroll;
12-15 to four-coil scroll; curves 16-19 to the case of five and more layered
structure of the nanoribbon scroll. The dashed lines are the fitting
curves: (a) $N_c=0.33L^{0.63}$, (b) $R_1=0.67L^{0.2}$, $R_2=0.4L^{0.47}$,
and (c) $\omega_1=36/L$, with $L$, $R_1$, and $R_2$ given in nanometers.
}
\label{fg08}
\end{figure}

Let us describe how the scroll structure and the lowest natural frequency depend
on the chain length (see Fig.~\ref{fg07} and Fig.~\ref{fg08}). Various conformations
can be naturally characterized by the number of coils $N_c$, energy per node $E_0$,
inner and outer radii $R_1$, $R_2$.

Single-coil configuration is the only possible in the chain length range
$5.772\le L\le 7.000$~nm [Fig.~\ref{fg07}~(a)]. Double coil configuration is stable
in case of $7.245\le L\le 20.508$~nm [Fig.~\ref{fg07}~(b), (c)]. The chain length
range $18.052\le L\le 36.718$~nm corresponds to stable three coil scrolls
[Fig.~\ref{fg07}~(d), (e)]. For the case of $33.771\le L\le 51.542$~nm the
four-coiled scrolls are observed [Fig.~\ref{fg07}~(f), (g)]. If nanoribbon length
$L\ge 51.0$~nm, the scrolls with five or more coils exist [Fig.~\ref{fg07}~(h)].

One can see that for nanoribbons with the same length two stable configurations
are possible. For example, in the length range $18.052\le L\le 20.508$~nm stable
two or three coil scrolls exist [Fig.~\ref{fg07}~(c) and (d)]. Nanoribbons of
length $33.771\le L\le 36.718$~nm can be  packed in three or four coil scrolls
[Fig.~\ref{fg07}~(e) and (f)]. The reason for bistability is the result of the
interaction of the nanoribbon ends. One stable configuration is when the ends
are close to each other and another one is for somewhat overlapped ends. The degree
of overlapping decreases with increasing nanoribbon length, as it can be seen
in Fig.~\ref{fg07}. Increase of the chain length leads to weakening of the
interaction between ends and thus to weakening of the bistability of the scroll packing.

Increase in the nanoribbon length $L$ results in the monotonous increase in the number of
coils $N_c$ according to the power law $N_c\approx 0.33L^{0.63}$,
see Fig.~\ref{fg08}~(a). The inner scroll radius $R_1$ increases much slower with
$L$ than the outer radius $R_2$: $R_1\approx 0.67L^{0.2}$, $R_2\approx 0.4L^{0.47}$,
see Fig.~\ref{fg08}~(b). Here $L$, $R_1$, and $R_2$ are given in nanometers.

The eigenmode having lowest positive frequency $\omega_1$ is the twisting-untwisting mode
when the atoms move along the Archimedes spiral. The second and the third lowest
eigenfrequencies correspond to lateral compression-extension of the scroll. The scroll
symmetry is lowered by the nanoribbon ends and for this reason the lateral compression
in the two orthogonal directions is characterized by the close (but not equal)
frequencies $\omega_2$ and $\omega_3$. These frequencies depend on $L$ nonmonotonically,
see Fig.~\ref{fg08}~(c). For $\omega_1$ the general trend is the reduction of
the frequency with the growth in $L$ according to the law $\omega_1\approx 36/L$ for $L\rightarrow\infty$. This is in line with the asymptotic behavior obtained
analytically in Refs. \cite{spcg09apl,spg10amss}.

\section{Frequency spectrum of the nanoribbon and nanoribbon scroll \label{sc5}}

Let us perform the full-atomic three-dimensional modeling of the dynamics of
the nanoribbon scrolls to verify the two-dimensional chain model.

\begin{figure}[t,b]
\begin{center}
\includegraphics[angle=0, width=1\linewidth]{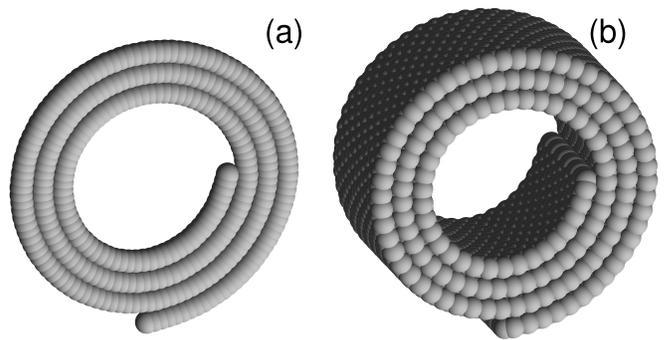}
\end{center}
\caption{
(a) Two dimensional chain model and (b) corresponding three dimensional full-atomic model of the nanoribbon scroll with the nanoribbon length $L_x=36.84$~nm
and width $L_z=2.55$~nm ($N=301$, $K=6$, number of atoms $N_{\rm all}=2NK=3612$).
}
\label{fg09}
\end{figure}

Let the set of two dimensional vectors $\{ {\bf u}_n^0=(u_{n,1}^0,u_{n,2}^0)\}_{n=1}^N$
is the solution of the minimization problem Eq. (\ref{f14}), describing a scroll packing
of the nanoribbon of length $L_x=(N-1)a$. For the nanoribbon of width $L_z=3Kr_0$
(the translational cell consist of $4K$ carbon atoms), then the coordinates of the atoms
in $n$-th cell are
\begin{eqnarray}
x_{n,4(k-1)+1}&=&u_{2n-1,1}^0,~y_{n,4(k-1)+1}=u_{2n-1,2}^0, \nonumber
\\ z_{n,4(k-1)+1}&=&3(k-1)r_0, \nonumber \\
x_{n,4(k-1)+2}&=&u_{2n,1}^0,~~~~y_{n,4(k-1)+2}=u_{2n,2}^0, \nonumber \\
z_{n,4(k-1)+1}&=&r_0/2+3(k-1)r_0,\nonumber\\
x_{n,4(k-1)+3}&=&u_{2n,1}^0,~~~~y_{n,4(k-1)+3}=u_{2n,2}^0, \label{f20} \\
z_{n,4(k-1)+1}&=&3r_0/2+3(k-1)r_0,\nonumber\\
x_{n,4(k-1)+4}&=&u_{2n-1,1}^0,~y_{n,4(k-1)+4}=u_{2n-1,2}^0, \nonumber \\
z_{n,4(k-1)+1}&=&2r_0+3(k-1)r_0, \nonumber\\
k&=&1,2,...,K,~~n=1,2,...,N/2, \nonumber
\end{eqnarray}
where, as above, $r_0$ is equilibrium C-C valence bond length. In the case of odd
number of cells, $N$, the nanoribbon consists of $N_{\rm all}=2NK$ carbon atoms.
The two dimensional chain model and the full-atomic nanoribbon scroll are
presented in Fig.~\ref{fg09} for the nanoribbon length $L_x=36.84$~nm
and width $L_z=2.55$~nm ($N=301$, $K=6$, number of atoms $N_{all}=2NK=3612$).

A set of interaction potentials (\ref{fp1}), (\ref{fp2}), (\ref{fp3}), (\ref{f10}) was used
for modeling of the nanoribbon dynamics. Valence bonds between neighboring atoms in the graphene
plane are described by the Morse potential (\ref{fp1}), valence and torsional angles by the
potentials (\ref{fp2}) and (\ref{fp3}). Weak van der Waals interactions between scroll coils
are described by the Lennard-Jones potential (\ref{f10}). Let us consider the edge carbon
atoms chemically modified by hydrogen atoms and thus, having mass of $M_0=13m_p$, under the
assumption that the interaction potentials for edge and internal atoms are the same.

Dynamics of the nanoribbon having size $L_x=(N-1)a\times 3Kr_0$ is described by the Langevin
equations
\begin{eqnarray}
M_{n,l}\ddot{\bf u}_{n,l}=-\frac{\partial H}{\partial {\bf u}_{n,l}}-\Gamma M_{n,l}\dot{\bf u}_{n,l}+\Xi_{n,l},\label{f21}\\
n=1,2,...,N,~~l=1,2,...,4K, \nonumber
\end{eqnarray}
where ${\bf u}_{n,l}=(x_{n,l},y_{n,l},z_{n,l})$ is the three dimensional radius-vector
of $(n,l)$-th atom, $M_{n,l}$ is the atom mass ($M_{n,l}=12m_p$ for the internal atoms
and $M_{n,l}=13m_p$ for the edge atoms). Here $H$ is the nanoribbon Hamiltonian,
$\Gamma=1/t_r$ is the friction coefficient, (velocity relaxation time is $t_r=0.4$~ps),
random forces vectors $\Xi_{n,l}=(\xi_{n,l,1},\xi_{n,l,2},\xi_{n,l,3})$ are normalized
as follows
$$
\langle\xi_{n,l,i}(t_1)\xi_{m,k,j}(t_2)\rangle=2Mk_BT\delta_{nm}\delta_{lk}\delta_{ij}\delta(t_1-t_2).
$$

The set of equations of motion Eq. (\ref{f21}) is integrated numerically. Initial
conditions corresponding to the stationary scroll packing of the two-dimensional chain
model Eq. (\ref{f20}) were used. We took $K=6$ and $N=101$, 201, 301, and 401 to simulate
the nanoribbon of width $L_z=3Kr_0=2.5524$~nm and length $L_x=(N-1)a=12.28$, 24.56, 36.84,
49.12~nm. Numerical simulation at low temperature $T=3$~K has shown the absence of any
noticeable changes in the initial structure of the scroll, i.e. the stationary configuration
of the three dimensional scroll is well described by the equilibrium configuration of the
two-dimensional chain. This confirms the high accuracy of the two-dimensional chain model.

\begin{figure}[t,b,p]
\begin{center}
\includegraphics[angle=0, width=1\linewidth]{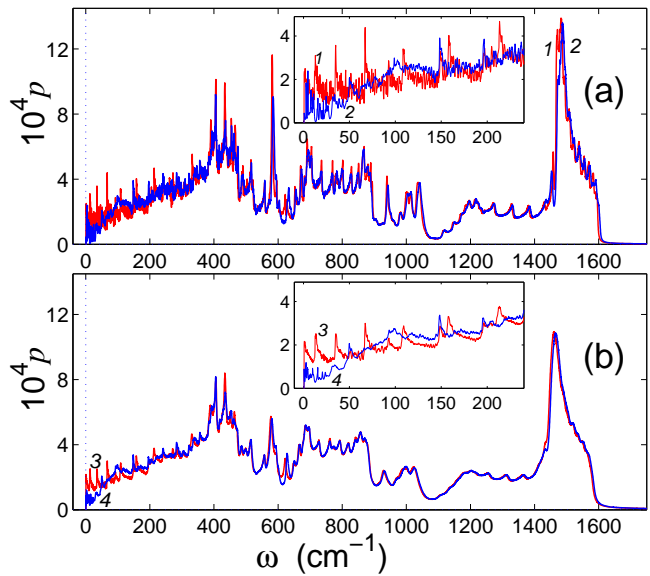}
\end{center}
\caption{
Phonon density of states $p(\omega)$ calculated from thermal vibrations of
atoms for the flat nanoribbon (red curves 1 and 3) and for the scroll (blue curves 2
and 4) at temperatures (a) $T=300$K and (b) $T=900$K. The nanoribbon length and width are
$L_x=36.84$~nm and $L_z=2.55$~nm, respectively.
}
\label{fg10}
\end{figure}

In order to find the phonon density of states for the full-atomic flat nanoribbon and for the
full-atomic scroll, the Langevin equations Eq. (\ref{f21}) were integrated for $10$~ps
to achieve the state of thermal equilibrium at the desired temperature, and then the
thermostat was switched off and free dynamics of atoms was studied. It was demonstrated
numerically that the flat and scrolled nanoribbons of length $L_x=36.84$~nm are both stable
in the temperature range $0\le T\le 900$~K. The use of full-atomic model allows us to find
the time dependence of the particle velocity on time $\dot{\bf u}_{n,l}(t)$ and then the
density of phonon states $p(\omega)$ normalized such that $\int_0^\infty p(\omega)d\omega=1$.
The density of phonon states was determined from 600 homogenously distributed atoms and 256
independent realizations of the initial thermalized nanoribbon state in order to increase
the calculation accuracy. The result for the nanoribbon of length $L_x=36.84$~nm and width
$L_z=2.55$~nm is shown in Fig.~\ref{fg10} for the flat nanoribbon (red curves 1 and 3) and
for the scroll (blue curves 2 and 4) at temperatures (a) $T=300$K and (b) $T=900$K.
As one can see, the frequency spectra of the flat and scrolled nanoribbons are very close.
Certain difference can be observed only in the low $\omega <150$ cm$^{-1}$ and high
$\omega>1450$ cm$^{-1}$ frequency intervals. In the range $\omega<50$ cm$^{-1}$ the scroll
has phonon density more than two times smaller than the flat nanoribbon. This is due to
the fact that the rigidity of scroll is higher than that of nanoribbon and the low-frequency
bending and torsional vibration modes are absent in the scroll. At high frequencies
a small blue shift (by 5 cm$^{-1}$) of the oscillation frequencies is observed for the scroll.
Scrolling of the nanoribbon leads to a moderate increase of oscillations frequencies in
the range $\omega> 1450$ cm$^{-1}$ due to the van der Waals interactions of atoms belonging to
adjacent layers of the scroll.

\section{Thermal expansion of scrolls \label{sc6}}

The scroll structure is stabilized by the weak van der Waals bonds acting between coils.
Thermal fluctuations weaken such bonding leading to partial untwisting or full opening
of the scroll. Fully opened scroll transforms to the flat GNR, while partial untwisting
results in a reduction of the number of coils and in a growth of the scroll diameter.

The full-atomic model does not allow to simulate the long-term dynamics of wide
multi-coiled nanoribbon scrolls due to the computer capacity limitations. For example,
thermal expansion of the scroll can hardly be treated by the full-atomic model
and this problem is addressed here in frame of the two-dimensional chain model.

For the simulation of thermal vibrations of the chain the Langevin equations
were used
\begin{eqnarray}
M\ddot{\bf u}_n=-\frac{\partial H}{\partial {\bf u}_n}-\Gamma M\dot{\bf u}_n+\Xi_n, \label{f22} \\
n=1,2,...,N, \nonumber
\end{eqnarray}
where ${\bf u}_n=(x_n,y_n)$ is the radius-vector of $n$-th node, $H$ is the
Hamiltonian of the chain Eq. (\ref{f1}), $N$ is the number of nodes in the chain,
$\Gamma=1/t_r$ is the friction coefficient (velocity relaxation time is $t_r=5$~ps), and
$\Xi_n=(\xi_{n,1},\xi_{n,2})$ is the two dimensional vector of normally distributed
random forces, normalized as
$$
\langle\xi_{n,i}(t_1)\xi_{m,j}(t_2)\rangle=2Mk_BT\delta_{nm}\delta_{ij}\delta(t_1-t_2)
$$
(here $k_B$ is the Boltzmann constant).

The set of equations of motion Eq.~(\ref{f22}) was  integrated numerically. The
stationary state of the scroll was used as an initial configuration.

Thermal stability of the scroll depends on the nanoribbon length $L$. The longer
is the nanoribbon the larger is the energy of van der Waals bonds per atom and the
higher is the thermostability. From simulations, it was found that single-coiled
scroll of the nanoribbon having $L=(N-1)a=12.28$~nm (number of nodes $N=101$) is
stable only for $T\le 330$K and at higher temperatures it fully opens in less than
$t=10$~ns. The two-coil scroll with $L=24.56$~nm $(N=201)$ was found to be stable
within the whole studied temperature range $T\le 960$K. Only partial untwisting
is observed for the scrolls with this and higher values of $L$ in this temperature
range.

\begin{figure}[t,b,p]
\begin{center}
\includegraphics[angle=0, width=1\linewidth]{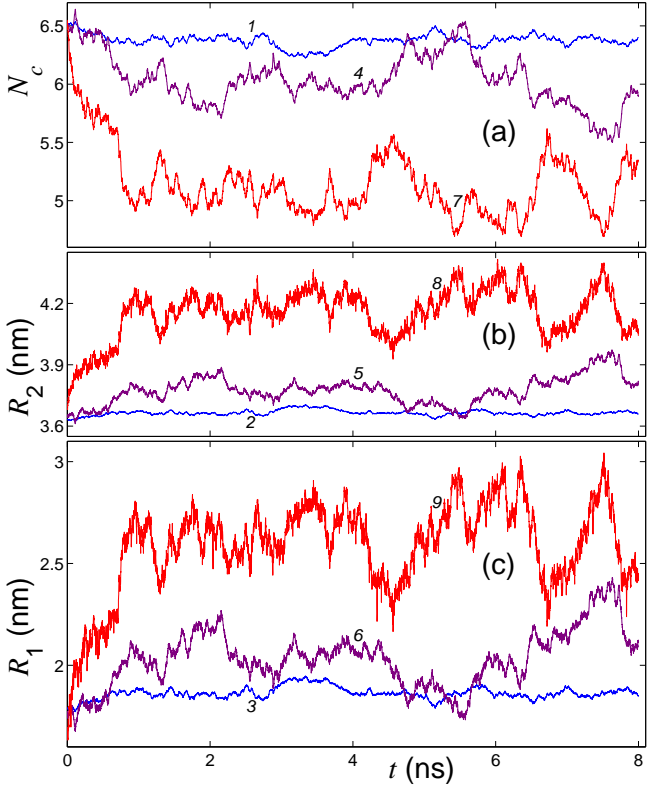}
\end{center}
\caption{
Time evolution of (a) number of coils $N_c$, (b) outer radius $R_2$, and
(c) inner radius $R_1$ for the scroll of the nanoribbon having length
$L=110.40$~nm, evaluated at the temperatures $T=30$~K (curves 1, 2, and 3),
$T=300$~K (curves 4, 5, and 6), and $T=900$~K (curves 7, 8, and 9).
}
\label{fg11}
\end{figure}

Let us take the scroll of the nanoribbon of length $L=110.40$~nm (number of nodes
$N=900$) for the study of its dynamics in the range of temperature $30\le T\le 960$K.
The dependence of number of coils $N_c$ and inner and outer radii $R_1$, $R_2$ of the scroll
on time is shown in Fig.~\ref{fg11}. As one can see from the graph, thermal vibrations
lead to decrease of the number of coils $N_c$ and growth of the radii $R_1$ and $R_2$.

\begin{figure}[t,b,p]
\begin{center}
\includegraphics[angle=0, width=1\linewidth]{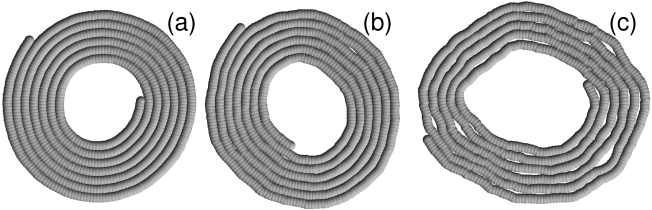}
\end{center}
\caption{
Typical scroll configurations for the nanoribbon having length
$L=110.40$~nm (number of chain nodes $N$= 900) at temperatures (a) $T=30$K,
(b) $T=300$K, and (c) $T=900$K.
}
\label{fg12}
\end{figure}

Typical scroll configurations at different temperatures are shown in Fig. \ref{fg12}.
At elevated temperatures the spiral structure is maintained and the inner and
outer radii of the scroll increase with temperature.

\begin{figure}[t,b,p]
\begin{center}
\includegraphics[angle=0, width=1\linewidth]{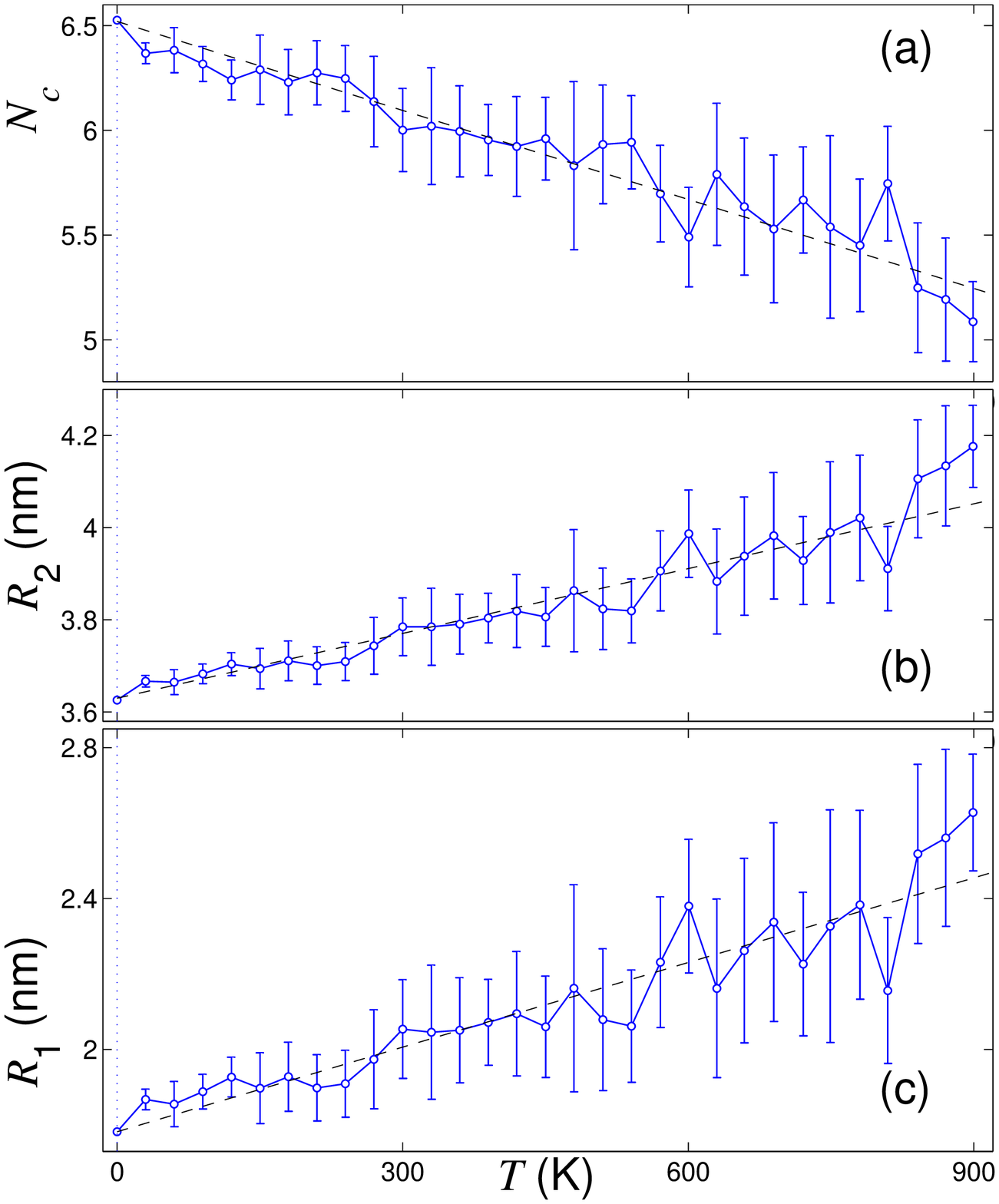}
\end{center}
\caption{
The temperature dependence of (a) the average number of coils $N_c$,
(b) the outer radius $R_2$, and (c) the inner radius $R_1$ of the scroll of the
nanoribbon having length $L=110.40$~nm. Dotted lines present the linear approximations.
}
\label{fg13}
\end{figure}

In about 1~ns thermalization of scroll is complete and it obtains its
equilibrium configuration. Further integration of the equations of atomic motion
allows to determine the averaged values of the coil number $\bar{N}_c$ and
inner and outer radii of the scroll, $\bar{R}_1$ and $\bar{R_2}$, corresponding
to the given temperature. This values, as the functions of temperature, are plotted
in Fig.~\ref{fg13}. It can be seen that the number of coils decreases linearly,
while the radii demonstrate a linear increase with temperature as $\bar{N}_c(T)\approx 6.52-0.00142T$,
$\bar{R}_i(T)\approx R_i^0+c_1T$, where $i=1,2$, $R_1^0=1.782$~nm, $R_2^0=3.63$~nm,
$c_1=0.00075$~nm/K, and $c_2=0.00047$~nm/K. The relative increase in the outer
radius of the scroll of the nanoribbon having length $L=110.40$~nm is
$\bar{R}_2(T)/R_2^0\approx 1+cT$, where the coefficient of linear thermal expansion
is equal to $c=c_2/R_2^0=1.3\cdot 10^{-4}$K$^{-1}$. It was found that $c$ depends on
$L$ such that $c$ is higher for smaller $L$. For example, for $L=73.56$~nm one has
$c=1.4\cdot 10^{-4}$K$^{-1}$; for $L=36.84$~nm $c=1.8\cdot 10^{-4}$K$^{-1}$; and
for $L=25.56$~nm $c=3.7\cdot 10^{-4}$K$^{-1}$.

Note that the coefficient of linear thermal expansion calculated for the graphene scroll
outer radius is two (one) orders of magnitude larger than that for graphite in $a$
($c$) direction (see \cite{CTE} and references therein reporting on the experimental
data) and two orders of magnitude larger than that for diamond \cite{CTEdiamond}.

\section{Conclusions \label{sc7}}

In this paper, the two-dimensional chain model (see Fig.~\ref{fg02}) was developed
to accurately and effectively describe the dynamics of folded and rolled conformations
of graphene nanoribbons. The Hamiltonian of the model is given by Eq.~(\ref{f1}) and
it takes into account the tensile and bending rigidity of the nanoribbon, as well as
the van der Waals interactions between layers of the nanoribbon. The chain model
describes only such modes of nanoribbon deformation [see Fig.~\ref{fg01}~(a)], where
the atomic rows parallel to $z$-axis move as rigid units only in the $xy$-plane but not
in $z$-direction. The nanoribbon width effect is not taken into account. Parameters
of the chain model were fitted to reproduce the low-frequency part of the phonon
dispersion curves of the flat graphene nanoribbon (see Fig.~\ref{fg03}). The van der
Waals interactions were fitted by the modified Lennard-Jones potentials (see Fig.~\ref{fg04}).

The validity of the chain model was demonstrated by comparison of the structure
of the stationary nanoribbon scrolls with the results of full-atomic simulations.

Potential energy per atom was calculated for flat, rolled, double-folded, triple-folded,
and rolled-collapsed conformations of nanoribbon as the function of its length $L$ (see
Fig.~\ref{fg05} and Fig.~\ref{fg06}). It was found that in the range $L<5.77$~nm the
planar structure is the only stable configuration of the nanoribbon. For $L\ge5.77$~nm,
stable rolled structures exist. Chains with $L\ge 6.02$~nm ($L\ge 10.19$~nm) can support
stable double-folded (triple-folded) structures. The rolled-collapsed structure requires
the chain length $L\ge 10.19$~nm. The flat nanoribbon has the lowest energy for
$L< 10.93$~nm. For the nanoribbon length in the range $10.93 \le L<13.39$~nm the lowest
energy is observed for the double-folded configuration, and for $L\ge 13.39$~nm
the most energetically favorable is the rolled structure (nanoribbon scroll). These
results can be easily interpreted in terms of the competition between energy release
due to the formation of van der Waals bonds and energy absorption due to bending
of the nanoribbon.

Since GNR having length $L\ge 13.39$~nm have the lowest energy in the rolled
conformation among the other studied configurations, the nanoribbon scrolls
were studied in detail (see Fig.~\ref{fg07} and Fig.~\ref{fg08}). It was found that
single-coil configuration is the only possible in the chain length range
$5.772\le L\le 7.000$~nm. Double coil configuration is stable
for $7.245\le L\le 20.508$~nm. The chain length range
$18.052\le L\le 36.718$~nm corresponds to stable three coil scrolls. For the
case of $33.771\le L\le 51.542$~nm the four-coiled scrolls are observed.
For nanoribbon length $L\ge 51.0$~nm, the scrolls with five or more coils exist.
Increase in the nanoribbon length $L$ results in the monotonous increase in the number of
coils $N_c$ according to the power law $N_c\approx 0.33L^{0.63}$. The inner scroll
radius $R_1$ increases much slower with $L$ than the outer radius $R_2$:
$R_1\approx 0.67L^{0.2}$, $R_2\approx 0.4L^{0.47}$. Here $L$, $R_1$, and $R_2$
are given in nanometers. The twisting-untwisting eigenmode having lowest positive
frequency $\omega_1$ was calculated as the function of nanoribbon length. For long
nanoribbons the asymptotic law was found
$\omega_1\approx 36/L$ for $L\rightarrow\infty$, which is in line with the earlier
theoretical studies \cite{spcg09apl,spg10amss}.

Full-atomic model was used to calculate phonon density of states for flat
and scrolled nanoribbons of length $L_x=36.84$~nm and width $L_z=2.55$~nm at
300 and 900~K (see Fig.~\ref{fg10}). It was shown that the phonon spectra
for the two conformations are very close in the entire frequency range.
Small difference can be observed only in the low $\omega <150$ cm$^{-1}$ and high
$\omega>1450$ cm$^{-1}$ frequency intervals. This can be explained by the higher
rigidity of scroll due to formation of the van der Waals bonds between adjacent
layers of the scroll. The low-frequency bending and torsional vibration modes
are absent in the scroll.

One of the most important findings of the present study has emerged from the
application of the developed chain model to the simulation of the long-term
dynamics of nanoribbon scrolls at different temperatures. It was found that
the relative increase in the outer radius of the scroll of the nanoribbon
having length $L=110.40$~nm is characterised by the coefficient of
linear thermal expansion of $c=1.3\cdot 10^{-4}$K$^{-1}$.
It was found that $c$ depends on $L$ such that $c$ is higher for smaller $L$.
For example, for $L=73.56$~nm one has $c=1.4\cdot 10^{-4}$K$^{-1}$; for
$L=36.84$~nm $c=1.8\cdot 10^{-4}$K$^{-1}$; and for $L=25.56$~nm
$c=3.7\cdot 10^{-4}$K$^{-1}$. The coefficient of linear thermal expansion
calculated for the graphene scroll outer radius is two (one) orders of
magnitude larger than that for graphite in $a$ ($c$) direction \cite{CTE}
and two orders of magnitude larger than that for diamond \cite{CTEdiamond}.
Such anomaly in the coefficient of thermal expansion can be used in the design
of nanosensors or other nanodevices.

The developed planar chain model can help to address problems related to
the dynamics of open graphene structures not treatable by the full-atomic
simulations. The results obtained in frame of the chain model could provide
a better understanding of the mechanical properties of CNS-based nanodevices.

\section{Acknowledgements}
A.V.~Savin thanks financial support provided by the Russian Science
Foundation, grant N 14-13-00982, and the Joint Supercomputer Center
of the Russian Academy of Sciences for the use of computer facilities.
E.A.~Korznikova is grateful for the financial support from the
President Grant for young scientists (grant N MK-5283.2015.2).
S.V.~Dmitriev appreciates the support from the Tomsk State University
Academic D.I.~Mendeleev Fund Program.

\end{document}